\documentclass[conference]{IEEEtran}
\IEEEoverridecommandlockouts
\usepackage{cite}
\usepackage{amsmath,amssymb,amsfonts}
\usepackage{algorithm}
\usepackage[noend]{algpseudocode}
\usepackage{multirow}
\usepackage{multicol}
\usepackage{scalerel}
\usepackage{subcaption}

\usepackage{graphicx}
\usepackage{textcomp}
\usepackage{xcolor}
\usepackage{url}
\usepackage{balance}

\makeatletter
\def\BState{\State\hskip-\ALG@thistlm}
\makeatother

\newcommand\blfootnote[1]{%
  \begingroup
  \renewcommand\thefootnote{}\footnote{#1}%
  \addtocounter{footnote}{-1}%
  \endgroup
}

\usepackage[normalem]{ulem}
\def\BibTeX{{\rm B\kern-.05em{\sc i\kern-.025em b}\kern-.08em
    T\kern-.1667em\lower.7ex\hbox{E}\kern-.125emX}}
\begin{document}


\title{Fast Outage Analysis of Large-scale Production Clouds with Service Correlation Mining}

\author{\IEEEauthorblockN{
Yaohui Wang\IEEEauthorrefmark{1}\IEEEauthorrefmark{6},
Guozheng Li\IEEEauthorrefmark{2},
Zijian Wang\IEEEauthorrefmark{1}\IEEEauthorrefmark{6},
Yu Kang\IEEEauthorrefmark{3},
Yangfan Zhou\IEEEauthorrefmark{1}\IEEEauthorrefmark{6},
Hongyu Zhang\IEEEauthorrefmark{4},
Feng Gao\IEEEauthorrefmark{5},\\
Jeffrey Sun\IEEEauthorrefmark{5},
Li Yang\IEEEauthorrefmark{5},
Pochian Lee\IEEEauthorrefmark{5},
Zhangwei Xu\IEEEauthorrefmark{5},
Pu Zhao\IEEEauthorrefmark{3},
Bo Qiao\IEEEauthorrefmark{3},
Liqun Li\IEEEauthorrefmark{3},
Xu Zhang\IEEEauthorrefmark{3},
Qingwei Lin\IEEEauthorrefmark{3}
}

\IEEEauthorblockA{
\IEEEauthorrefmark{1}School of Computer Science, Fudan University, China\\
\IEEEauthorrefmark{6}Shanghai Key Laboratory of Intelligent Information Processing, China\\
\IEEEauthorrefmark{2}School of Electronics Engineering and Computer Science, Peking University, Beijing, China\\
\IEEEauthorrefmark{3}Microsoft Research, Beijing, China\\
\IEEEauthorrefmark{4}School of Electrical Engineering and Computing, The University of Newcastle, Australia\\
\IEEEauthorrefmark{5}Microsoft Azure, Redmond, USA\\
}}

\maketitle

\begin{abstract}

Cloud-based services are surging into popularity in recent years. However, outages, \textit{i.e.,} severe incidents that always impact multiple services, can dramatically affect user experience and incur severe economic losses. Locating the root-cause service, \textit{i.e.}, the service that contains the root cause of the outage, is a crucial step to mitigate the impact of the outage. In current industrial practice, this is generally performed in a bootstrap manner and largely depends on human efforts: the service that directly causes the outage is identified first, and the suspected root cause 
is traced back manually from service to service during diagnosis until the actual root cause is found. 
Unfortunately, production cloud systems typically contain a large number of interdependent services. Such a manual root cause analysis is often time-consuming and labor-intensive. In this work, we propose COT, 
the first outage triage approach that considers the global view of service correlations. 
COT mines the correlations among services from outage diagnosis data. 
After learning from  historical outages, COT can infer the root cause of emerging ones accurately. We implement COT and evaluate it on a real-world dataset containing one year of data collected from Microsoft Azure, one of the representative cloud computing platforms in the world. Our experimental results show that COT can reach a triage accuracy of 82.1\%$\sim$83.5\%, which outperforms the state-of-the-art triage approach by 28.0\%$\sim$29.7\%.
\blfootnote{This work was done at Microsoft Research (Beijing, China). Yaohui Wang and Guozheng Li contribute equally to this work.}

\end{abstract}


\section{Introduction}\label{introduction}

Cloud computing has become increasingly popular in recent years. Many companies have migrated their services to various cloud computing platforms, {\em e.g.}, Microsoft Azure, Amazon AWS, and Google Cloud. These platforms provide a variety of services to millions of users from all over the world every day. 
Availability
is one of the most critical concern to cloud computing platforms, influencing the user experience and the cloud providers' revenue significantly.

Although tremendous efforts have been devoted to maintaining high service availability~\cite{lin2018Predicting, ameller2016survey, Evaluating2013Li, tilley2012software}, cloud computing platforms still encounter many incidents, \textit{i.e.}, unplanned interruptions of the services. These incidents,
especially outages ({\em i.e.}, a group of related severe incidents that may impact multiple services), cause  significant economic losses. According to a study conducted on 12.4 million US  
businesses\footnote{\url{https://www.lloyds.com/news-and-insights/risk-reports/library/cloud-down}}, a failure that takes a top cloud provider offline in the US for 3 to 6 days would result in \$15 billion of economic loss. In this regard, once an outage occurs, it should be mitigated in a timely manner to minimize its impact. One important step of such outage mitigation is locating its root-cause service, which allows the corresponding responsible team to fix the outage. 

In current industrial practice, outages are typically declared on an originating service, {\em i.e.}, the service where the outage manifests. The team responsible for this service will analyze the outage, and may redirect the outage to the team of another service, until it eventually reaches the root-cause service where the outage can be fixed.  
Our empirical study (discussed in Section~\ref{preliminary}) on a representative large-scale cloud shows that assigning an outage to wrong services leads to a time-consuming root cause analysis process.

Unfortunately, fast and accurate root cause analysis of outage is a very challenging task.
Firstly, the number of services is vast in large-scale cloud computing platforms. The dependencies among the services are incredibly sophisticated. Many services have interdependencies (\textit{e.g.}, the \textit{micro-service management service} is used to deploy the 
\textit{resource management service}, 
while the computing node where the \textit{micro-service management service} is deployed is managed by the \textit{resource management service}). Many dynamic dependencies are even implicit for engineers (\textit{e.g.}, asynchronous communication, virtual routers, virtual disks), and some services deployed on the same node may affect each other (\textit{e.g.}, monitoring services and functional services). When an outage occurs, massive noisy alerts 
might be reported, usually as incident tickets, due to the notorious flooding alarm problem in cloud computing platforms~\cite{chen2020towards}. As a result, it is difficult to decide the root-cause service. According to our empirical study, the originating service and root-cause service are different for over 52\% outages.
Secondly, during the root cause analysis process, the engineers of one service
typically have only a partial view of the outage, since they may only be familiar with their own service and its closely-related ones. 
Consequently, the outage may be passed from one service team to another, resulting in a long 
team assignment chain for root cause analysis. 
Our empirical study shows that such a long chain dramatically prolongs the  time required for root cause analysis.

Incident triage, which aims at locating the root cause of an incident in a small scope of services~\cite{chen2019empirical}, has been extensively investigated in the literature. For example, DeepCT~\cite{chen2019continuous} is a state-of-the-art incident triage method. However, 
such methods rely
only on the information of the incident {\em per se}, {\em e.g.}, the title and summary of incident report and engineers' discussions during diagnosis. They do not take the global view of 
service correlations 
into consideration. As a result, they are not suitable for root cause analysis of outage, since an outage typically involves many correlated services.   

In this paper, we propose COT (\textbf{C}orrelation-based \textbf{O}utage \textbf{T}riage) to facilitate fast and accurate outage root cause analysis. 
COT is a first attempt in the literature, to our knowledge, which provides a global view of service correlations for outage root cause analysis. 
COT collects all the incidents in the same region and the adjacent time range with the outage. Based on the historical outage diagnosis data, it finds the correlations among these incidents and builds an incident correlation graph. By mapping incidents to their owning services, 
COT can further obtain the service correlation graph, which indicates the anomaly propagation patterns among services. It then takes the features extracted from these graphs of historical outages with their root-cause services as training samples to train a machine learning model. This model can then be used to predict the root-cause service for emerging outages.

We evaluate COT in two ways. We first conduct a quantitative experiment to compare the efficiency and effectiveness of COT and DeepCT, the state-of-the-art approach. The result shows that COT can perform root cause analysis of an outage within one minute, and outperforms DeepCT by 
28.0$\sim$29.7\% in accuracy.
Second, we evaluate the usability of COT through 
a representative real-world case study. The results show that COT could facilitate engineers to understand 
outages and 
locate the root-cause services. 

The contributions of this work are summarized as follows. 

\begin{itemize}
    \item We conduct the first empirical study on the outage triage problem in large-scale production clouds. The results are based on  representative real-world cloud services 
    and can facilitate further follow-up research.
    \item We propose COT, a generic cross-service outage triage approach. It is a correlation-based approach which predicts the root-cause service at the early stage of an outage with high accuracy. 
    \item We comprehensively evaluate COT by comparing it with the state-of-the-art triage approach 
    and by conducting a case  study  of two real-world outages.
\end{itemize}

The rest of the paper is organized as follows: Section~\ref{preliminary} presents the results of an empirical study of outage triage. Section~\ref{approach} describes the overall design of the correlation-based outage triage approach, and Section~\ref{experiment} introduces the implementation details of COT and the experimental settings. In Section~\ref{evaluation},  we evaluate the performance and time efficiency of COT and compare it to the state-of-the-art incident triage method. In Section~\ref{case_study}, we conduct a case study of two real-world outages to show the effectivness of COT. Section~\ref{threats} discusses the threats to the validity of our work. Section~\ref{related} presents some related work. Section~\ref{conclusion} concludes the paper.


\section{An Empirical Study of Outage Triage  }
\label{preliminary}


This section presents the first empirical study of outage triage in real-world cloud computing platforms. The study is based on Microsoft Azure, one of the representative cloud computing platforms in the industry. In the following, we first introduce the life cycle of an outage in Microsoft Azure, especially the differences between outage triage and incident triage. Next, we analyze $N$ outages collected from Microsoft Azure, ranging from 2020-01-01 to 2020-07-31, and present an empirical investigation of the outage reassignment during the outage triage process. Due to the confidential policy of Microsoft Azure, we use variables to represent the sensitive data, instead of disclosing the specific figures.

\subsection{The Life Cycle of Outages}
\label{outage-life-cycle}

In cloud computing platforms, incidents/outages are the unplanned interruptions of the services, which could be caused by many factors, such as power failures, hardware failures, configuration problems, and code bugs. For each incident/outage, engineers assign a severity level based on their potential impact on users. The severity level ranges from 0 to 4, where incidents with severity 0 have the highest priority and may bring negative impacts to customers, and severity 4 incidents are least important and do not need to be dealt with immediately. Engineers only declare a small portion of incidents as outages, which are severe incidents and often involve multiple services in the cloud computing platform. 

\begin{figure}
\centering
    \includegraphics[width=0.45\textwidth]{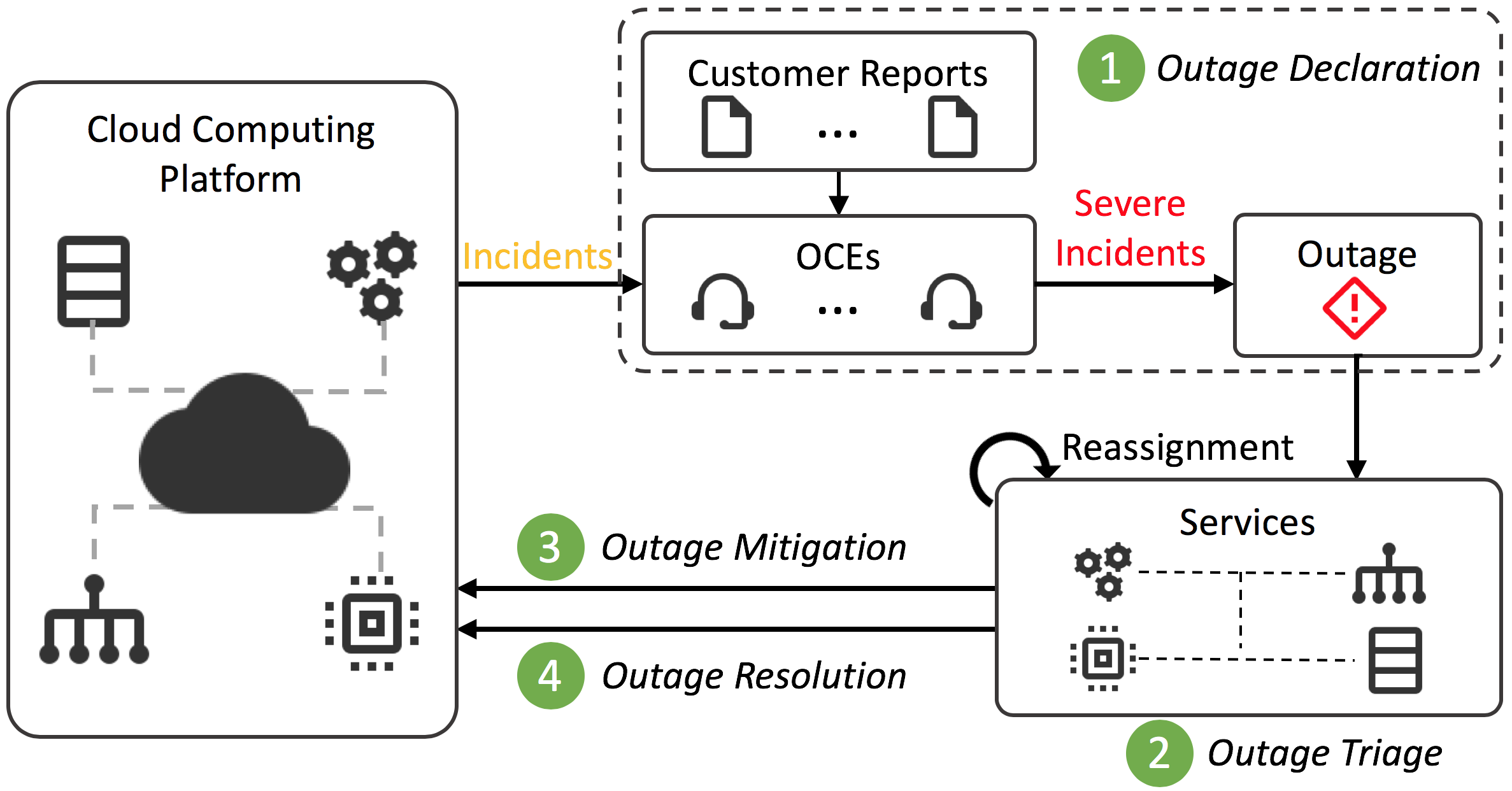}
    \caption{The life cycle of an outage.}
    \label{fig:outage-triage}
\end{figure}

Previous research has investigated four typical incident management procedures, including incident reporting, triage, mitigation, and resolution~\cite{chen2019outage}. As illustrated in Figure~\ref{fig:outage-triage}, the life cycle of outages is similar to that of incidents. The following mainly demonstrates the detailed practice of outage triage and emphasizes the differences between outages and incidents. 
After an incident is reported, On-call Engineers (OCEs) will declare it as outage if they find the incident may cause severe impact to end-users. 
Next, OCEs will try to restore the service as soon as possible through a set of pre-defined instructions, \textit{i.e.,} steps which tell engineer how to minimize the impact. These instructions are related to the incident which the outage is declared from. 
However, as many outages may have cross-service impacts (Section~\ref{outage-statistics}), the instructions usually do not work. For these outages, engineers need to do outage triage to locate the root-cause service and the responsible team.
During the outage triage process, after assigning the outage to a team, the engineers in this team will confirm whether or not they are responsible for the outage. If not, they will reassign the outage to another team which is potentially responsible. This process is called outage reassignment and it is repeated until the responsible team of the root-cause service is found.

After that, to contain the outage's impact to end-users, the responsible team will try to debug the service and use quick-fixes to mitigate the outage. Finally, to avoid the same failure from happening again, engineers need to study the outage in-depth and fix the problem permanently in the outage resolution phase. At the same time, to learn the lessons from the outage and assess its impact, engineers from all impacted services may dive into the details of the outage, find the related incidents, and mark the correlations among those incidents.

\begin{figure}
\centering
    \hfill
    \includegraphics[width=0.45\textwidth]{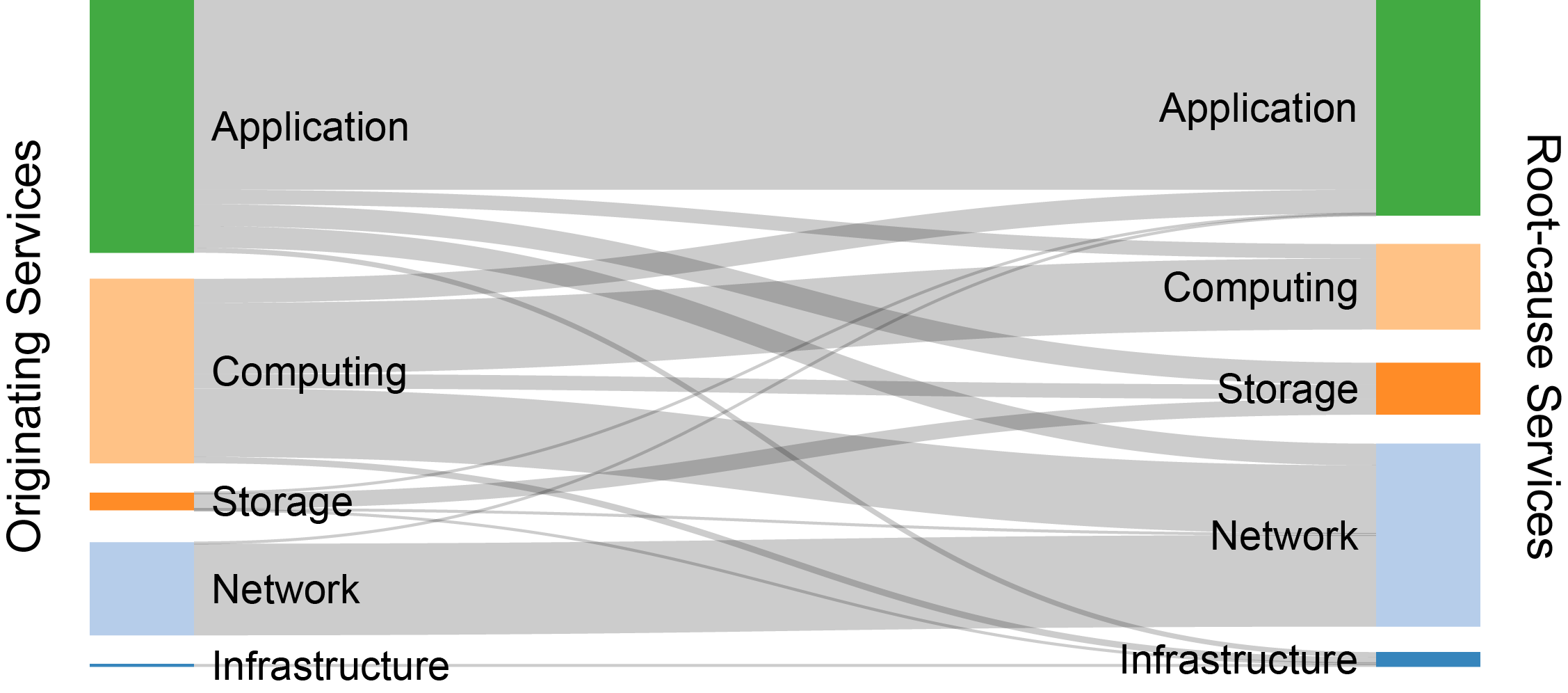}
    \hfill
    \caption{Outages' originating services and root-cause services.}
    \label{fig:originating-responsible}
\end{figure}

\subsection{An Empirical Analysis of Outage Triage}\label{outage-statistics}

Based on the outage data collected from Microsoft Azure, we conduct an empirical investigation of outage triage. The services in the online computing platform have complicated dependencies. To better understand the relationships among different services, we divide the services into five categories according to their functionality, namely \textit{Infrastructure}, \textit{Networking}, \textit{Storage}, \textit{Compute}, and \textit{Application}. 
Then we investigate the practice of outage triage from two aspects: the number and time cost of outage reassignments, respectively.

Services of different categories play various roles in the cloud computing platform. The \textit{Infrastructure} category lies on the lowest level of the cloud computing platform and is responsible for the hardware availability. All other services rely on the \textit{Infrastructure} category. The \textit{Network} category is responsible for network connectivity. The \textit{Storage} category is responsible for the management of storage resources. The \textit{Computing} category is responsible for the management of computing resources. The \textit{Application} category lies on the highest level of the cloud computing platform. Services in this category are exposed to customers and rely on the supporting services of other categories, directly or indirectly.

\begin{figure}
\centering
    \hfill
    \includegraphics[width=0.45\textwidth]{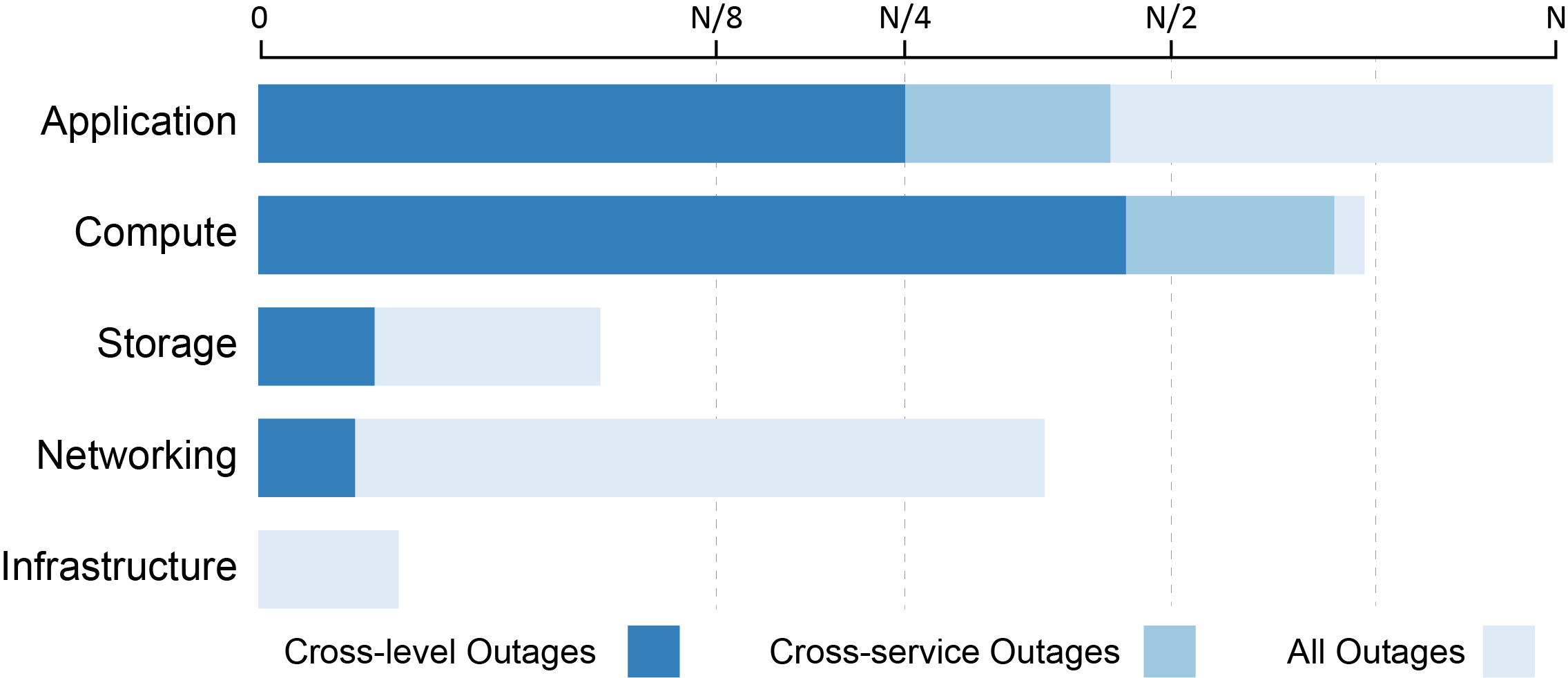}
    \hfill
    \caption{The amount of the cross-service and cross-level outages (sqrt scale).}
    \label{fig:cross-service-level}
\end{figure}

(1) \textbf{Number of Outage Reassignments.} The incident management system (IcM) records the reassignment behaviors among different services. Figure~\ref{fig:originating-responsible} shows the relationships between originating services and root-cause services of the collected outages. We can learn that the cross-level outages, whose originating service level and root-cause service level are different, exist in all categories except \textit{Infrastructure}. 
In particular, for the outages whose root-cause services belonging to the \textit{Networking} category, around 60\% are cross-level. 
Note that each category contains multiple services, and the outages which are not cross-level outages could still be cross-service outages. 
We categorize the outages according to their originating service level and compute the amount of the cross-level outages and cross-service outages. 
The results in Figure~\ref{fig:cross-service-level} show that for \textit{Storage}, \textit{Networking}, and \textit{Infrastructure} categories, nearly all the cross-service outages are also cross-level outages. 
For cross-service outages in the \textit{Application} and \textit{Compute} categories, about 60\% of them are cross-level.

We further analyze the average number of outage reassignments from different originating service categories. The result in Table~\ref{tab:avg-reassignment} shows that the average reassignment number of all outages is nearly one. In particular, the outages whose originating services are in the \textit{Compute} category have significantly higher average reassignment rate. This fact further motivates the necessity of accurate outage triage at the service level.

\begin{table}[htbp]
    \centering
    \caption{The average number of outage reassignments from different originating service categories.}
    \begin{tabular}{| c | c | c | c | c | c | c}
            \hline
            Level & Networking & Storage & Compute & Application & All  \\
            \hline
            Avg &  0.678 & 0.407 & 1.395 & 0.802 & 0.963 \\
            \hline
    \end{tabular}
    \label{tab:avg-reassignment}
\end{table}

Among all cross-level outages, around 14\% are transferred from low-level categories to high-level categories. We analyze those outages and summarize the following two reasons.

\begin{enumerate}
    \item \textbf{Method Caller Errors.} Services in high-level categories always call the functions of services in the low-level categories, but the root cause of outages may exist in the caller side beside the callee side. For example,  an \textit{Application} service query data through a function of a \textit{Storage} service, but wrong parameters lead to outages. These outages' originating services belong to the \textit{Storage} category, but the root-cause services are from the \textit{Application} category.
    \item \textbf{Symbiotic Errors.} The services of different categories could influence each other because they run on the same physical nodes. For example, a service about managing network resources is from the \textit{Networking} category, and a monitoring application also runs on the same node. Some network outages occur to the cloud computing platform because of monitors taking too many physical nodes' resources. These outages' originating services are from the \textit{Networking} category, but the root-cause service is the monitoring application. 
\end{enumerate}

\begin{figure}
\centering
    \hfill
    \includegraphics[width=0.45\textwidth]{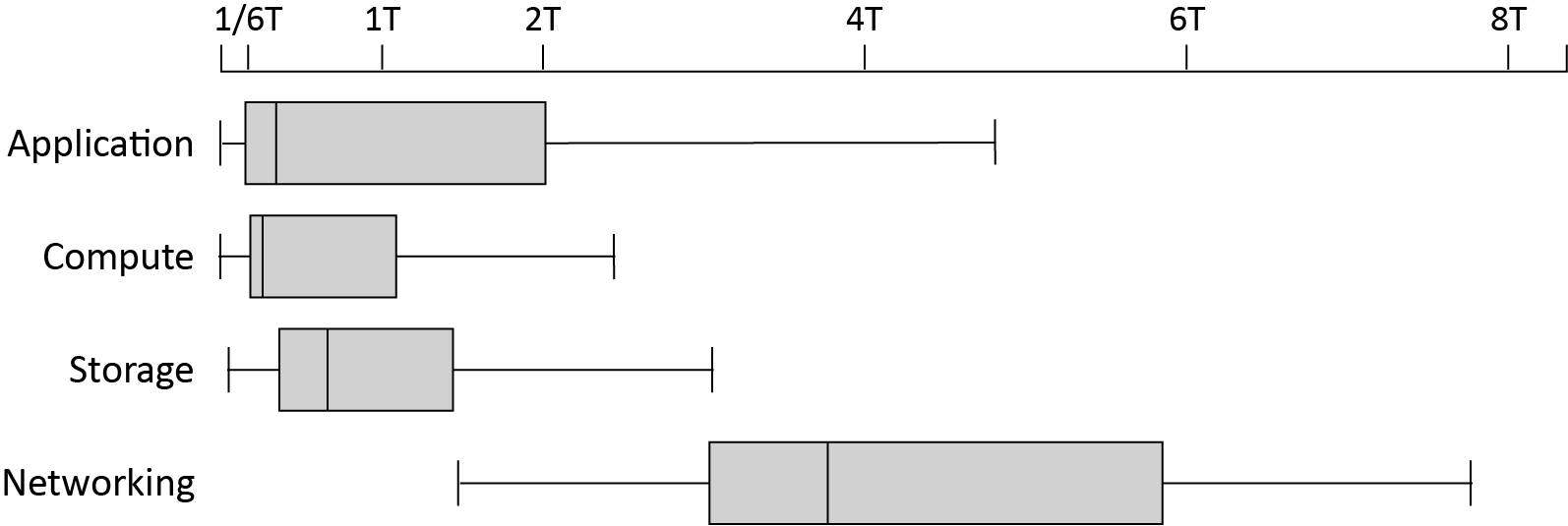}
    \hfill
    \caption{The outage reassignment time cost in different categories.}
    \label{fig:triage-time-cost}
\end{figure}

(2) \textbf{Outage Reassignment Cost.} The reassignment cost indicates the increment of the outage triage time because of assigning outages to the wrong services.
The IcM records an accurate time when transferring the outage to the specific services.
We calculate the time difference of outages transferring into and out of a wrong service as the cost of one outage reassignment. We can obtain the total reassignment cost of an outage by 
adding up all the reassignment cost of this outage.
We categorize the outages according to the originating service level and compute each category's average reassignment cost. The results in Figure~\ref{fig:triage-time-cost} demonstrate that the \textit{Networking} category has a relatively larger reassignment cost than other categories. 

Based on the above analysis of outage triage from the aspects of outage reassignment rate and cost, the outage reassignment at the service level could incur huge time cost. 
Accurate outage triage could significantly improve the efficiency of outage mitigation and reduce the cost.

\subsection{A Real-world Outage Triage Example}\label{preliminary_case}

\begin{figure}[ht!]
    \centering
    \includegraphics[width=0.49\textwidth]{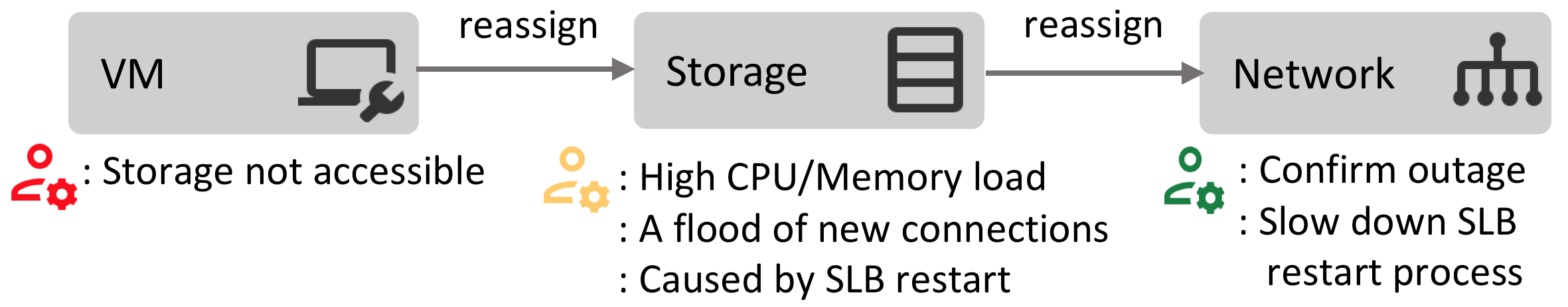}
    \caption{The triage path for the ``unexpected VM restart'' outage.}
    \label{fig:outage-triage-example}
\end{figure}

To help better understand the outage triage process, we show a real-world triage process for an ``unexpected VM restart'' outage. Figure~\ref{fig:outage-triage-example} shows the triage path for this outage. First, the VM (Virtual Machine) monitor detects the ``unexpected VM restart'' event on some VMs and reports an incident to the IcM. Engineers assess the incident at first. They consider that this incident may cause severe impact to end-users, 
and declare it as an outage. When diagnosing the \textit{VM} service, engineers find that the storage the VMs use is not accessible, and they transfer the outage to the related \textit{Storage} service. Then engineers of the \textit{Storage} service find that the CPU and Memory load of some storage services is unusually high, and the \textit{Storage} service is crashed due to an OOM (Out of Memory) exception. They also find the reason for this phenomenon is a flood of new connections to the storage cluster, which is caused by the restart of the SLB (Service Load Balancer). Finally, the outage's root cause is confirmed by the related \textit{Network} service, which is responsible for the SLB, and engineers in the \textit{Network} service mitigate this outage by slowing down the SLB restart process. According to this case, engineers transfer the outage from service to service and the engineers of each service need to understand the outage first, diagnose the service to confirm whether they are responsible for the outage and if not, they need to assign the outage to the engineers of another service manually.

\section{The Proposed Approach to Outage Triage}\label{approach}

\begin{figure*}[ht!]
    \centering 
    \includegraphics[width=0.9\textwidth]{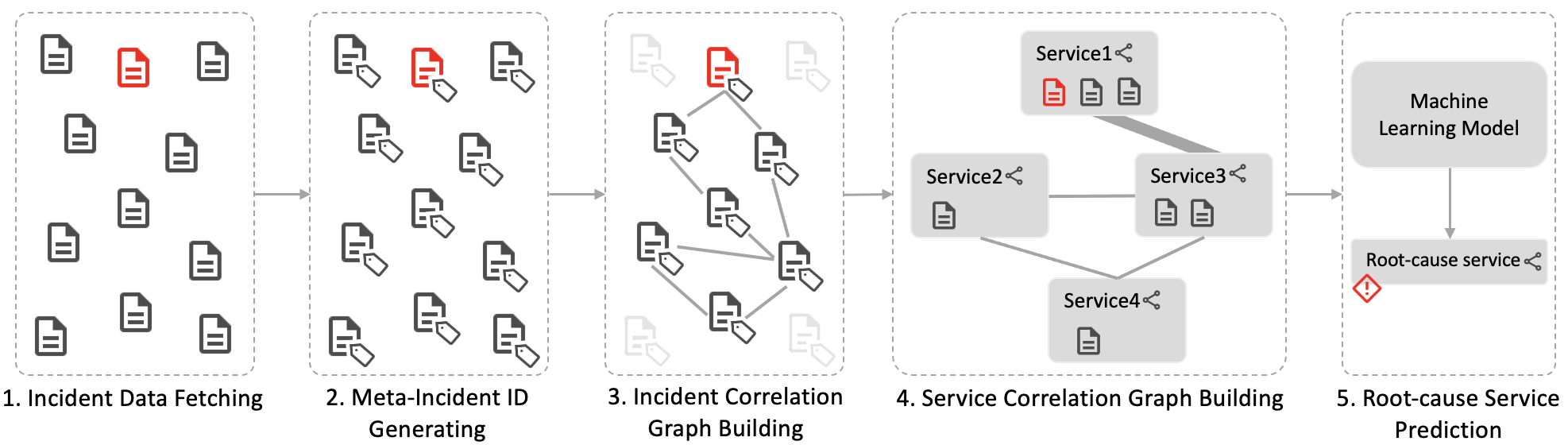}
    \caption{An Overview of COT.}
    \label{fig:approach_overview}
\end{figure*}

The example explained in Section \ref{preliminary_case}
shows that the service teams only know their direct dependency and always triage the outage following dependency relations. Therefore, locating the root-cause service without reassignment requires a global view of dependencies for all related services. 
It is not a trivial task for large-scale cloud computing platform because the dependencies of inner services are incredibly complicated and change frequently. 
Many services have interdependencies, many dependencies abstracted by the platform are even implicit for engineers, and some services deployed on the same node may affect each other.
Besides, service dependencies are not static but related to specific problems.

COT constructs the service correlation graph of an outage based on the incidents.
When a new outage occurs in the cloud computing platform, COT collects all the incidents in the same region and adjacent time range with the outage. 
Then it filters the incidents which are directly or indirectly related to the outage, and construct an incident correlation graph from these incidents. Finally it maps the incidents to services to construct a service correlation graph for the outage.
The outages caused by the same root-cause service have similar symptoms, \textit{i.e.,} service correlation graph (Section~\ref{case_study}), which inspires us to train a machine learning model to predict the root cause. 
Our model takes the service correlation graph of historical outages as training samples, and the label of each outage is the root-cause service, which is labeled by engineers after outage diagnosis. 

Figure~\ref{fig:approach_overview} gives an overview of our method, which contains five steps: incident data fetching, meta-incident ID generating, incident correlation graph building, service correlation graph building, and root-cause service prediction. In the following, we 
further explain the details of each step.

\subsection{Incident Data Fetching}

Since the impact of one outage usually has locality in time and space, we only fetch the incidents which are reported in the same region and adjacent time range with the outage. In Microsoft Azure, the medium time of finding the correct root-cause service is $T$. And our statistics on historical outages show that most of the early incidents related to an outage are reported within about $2T$ before the outage declaration time. So we set the start time of the time window to $2T$ before the outage declaration time, 
which allows us to capture most of the related early incidents. Also, to examine how our approach is better than the traditional outage triage process, the end time should be smaller than $T$. So we choose three time windows: $[-2T, \frac{1}{3}T]$, $[-2T, \frac{2}{3}T]$ and $[-2T, T]$. The last time window $[-2T, T]$ is to examine the performance of our approach when we get the same incident information as human engineers.

Note that the range of the time window should be different in different systems, as the statistics of the early outage related incidents and the average outage triage time are different. But the method to determine the time window should be general, thus people can follow this method to evaluate the performance of the approach using different time windows and choose the best time window that fits their system.

\subsection{Meta-Incident ID Generating}

To utilize the historical outage diagnosis data, we need to distinguish incidents by the abnormal symptoms they represent, \textit{i.e.,} ﬁnd past incidents that represent the same abnormal symptoms as newly occurring incidents. The ideal way is to use monitor IDs to identify different abnormal symptoms from incidents. This is reasonable for a well-designed monitoring system, 
where every monitor has a unique ID. But according to our experience of Microsoft Azure, the IcM system is a hub 
that gathers incidents reported by many different sources from hundreds, even thousands, of services. It is hard to define general rules to constrain how the monitors should report incidents. So, some monitors may use the same identification, some may monitor several different services at the same time, and some different monitors may monitor the same properties of the same resource type in different regions.

In Microsoft Azure, the title of an incident usually summarizes the symptom of the incident. However, we cannot directly use incident titles to distinguish different abnormal symptoms, since they are usually log-like texts and may contain variables like service metrics, time or location. Thanks to the well studied log parsing technology in recent years, we can use the log parsing method to extract incident report templates from their titles and use the templates to distinguish incidents by the abnormal symptoms they represent. We go through the historical incident data and assign a unique ID, called meta-incident ID, to each of the incident template to obtain an ``incident template - meta-incident ID'' mapping.

We follow the widely-used log-parsing approach SLCT~\cite{vaarandi2003data} to extract the template from incident titles automatically. The original SLCT~\cite{vaarandi2003data} log parser builds the word vocabulary over all the log texts, picks out candidate words that appear frequently, and uses these words to extract templates from log texts. But in our case, directly using SLCT to parse incident titles may bring one problem: some location representations may contain multiple words and numeric values, like \texttt{West US 2}, but the SLCT will treat each part of the location representation separately. To make the incident parsing more precise, we manually add the location representations in Microsoft Azure to the vocabulary.

\begin{figure}
    \centering
    \includegraphics[width=0.45\textwidth]{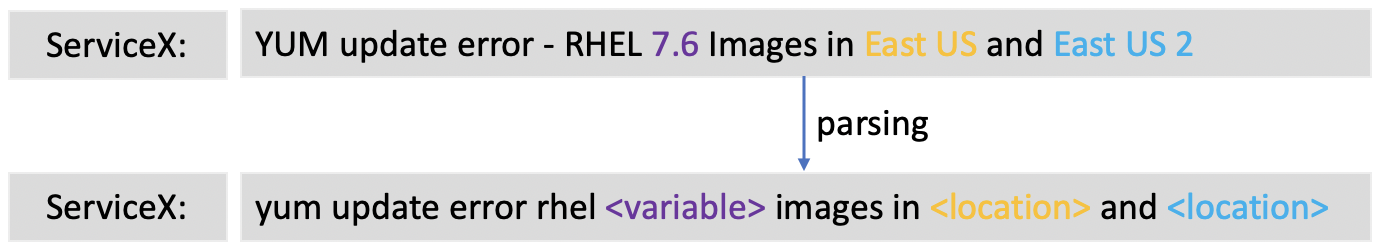}
    \caption{An example of parsing the template from an incident title.}
    \label{fig:incident-title-template}
\end{figure}

Figure~\ref{fig:incident-title-template} shows a log parsing example. The log parser first turns all words to lower-cases. The words in the built vocabulary stay unmodified, and words representing locations are recognized as \texttt{<location>} tokens, and others are recognized as \texttt{<variable>} tokens. Besides, to prevent the case that two different services use the same template, we combine the owning service with the template to identify the incidents.

For each incident in the past, we can use this method to extract its template. We consider each incident template as a different abnormal symptom in the system and give each of them a unique meta-incident ID. For a new incident, we can use the same incident-parsing method to extract the incident template and obtain its corresponding meta-incident ID.

\subsection{Incident Correlation Graph $G_I$}\label{incident_correlation_graph}

Since a large-scale cloud computing platform contains a large number of services, the reported incidents may come from many different services, and the amount is usually very large even within a small time window. Among them, only a small proportion of incidents are related to the outage (Section~\ref{case_study}). In current practice, finding related incidents of a new outage relies on heavy manual work of experienced engineers.

We use the historical outage diagnosis data to solve this problem. We first build the meta-incident ID correlation graph $G_M$ from historical data to infer the potential links for newly occurring incidents. For each pair of incidents labeled as correlated in the outage resolution phase (Section~\ref{outage-life-cycle}), we parse them and obtain a pair of meta-incident IDs. Then we take the meta-incident IDs as nodes and the their correlations as edges to build an meta-incident ID correlation graph $G_M$. $G_M$ is further used to build the incident correlation graphs for new outages. 

Now we can build the incident correlation graph $G_I$ as we already get the meta-incident IDs for the incidents fetched earlier and the correlations among the meta-incident IDs in $G_M$. Algorithm~\ref{alg:build_incident_correlation_graph} illustrates this process. Its input contains three parts: $G_M$ is the meta-incident ID correlation graph; $I$ is the set of incidents that are reported near the outage in time and locality; $I_O$ is the incident of the outage (\textit{i.e.,} the incident where the outage is declared from). Its output is the incident correlation graph $G_I$. $N_{G_I}$ and $E_{G_I}$ hold the incidents and edges in $G_I$, respectively. At the beginning of the procedure, $I_O$ is the only element in $N_{G_I}$, and $E_{G_I}$ is empty. Then for each loop, we go through each incident in $I$, get its meta-incident ID, and check whether there exist any links in $G_M$ that can link this incident to the existing graph. We will repeat this loop until no new incidents can be added to the graph.

\begin{algorithm}
\caption{The procedure of building $G_I$}
\label{alg:build_incident_correlation_graph}
\begin{flushleft}
    \hspace*{\algorithmicindent} \textbf{Input:} $G_M$, $I$, $I_O$\\
    \hspace*{\algorithmicindent} \textbf{Output:} $G_I$
\end{flushleft}
\begin{algorithmic}[1]
    \Procedure{BuildIncidentCorrelationGraph}{}
        \State $N_{G_I} \gets \{I_O\}$
        \State $E_{G_I} \gets \{\}$
        \State $N'_{G_I} \gets N_{G_I}$
        \State $E'_{G_I} \gets E_{G_I}$
    \BState \emph{loop}:
        \For {each incident $I_a$ in $I - N'_{G_I}$}
            \State $ID_a \gets$ meta-incident ID of $I_a$
            \For {each incident $I_b$ in $N'_{G_I}$}
                \State $ID_b \gets$ meta-incident ID of $I_b$
                \If {$<ID_a, ID_b>$ in $G_M$}
                    \State add incident $I_a$ to $N_{G_I}$
                    \State add edge $<I_a, I_b>$ to $E_{G_I}$
                \EndIf
            \EndFor
        \EndFor
        \If{size of $N'_{G_I}$ $\neq$ size of $N_{G_I}$}
            \State $N'_{G_I} \gets N_{G_I}$
            \State $E'_{G_I} \gets E_{G_I}$
            \State \textbf{goto} \emph{loop}
        \EndIf
        \State $G_I = \{N_{G_I}, E_{G_I}\}$
        \State \Return $G_I$
    \EndProcedure
\end{algorithmic}
\end{algorithm}

\subsection{Service Correlation Graph $G_S$}\label{approach_service_correlation_graph}

As we have mentioned in Section~\ref{outage-life-cycle}, every outage is well studied after mitigation, so the same root-cause bug of a service is unlikely to happen again in the future. Even for two outages that have the same root-cause service, their root-cause bugs are usually different, so as the abnormal metrics observed by monitors and the reported incidents. Thus it is hard to refer to historical incident patterns for new outages. However, although the malfunction of the root-cause service is usually caused by different bugs, the spreading paths of the anomaly among dependent services caused by the root-cause service have similar patterns (Section~\ref{case_study}). So we group the incidents in the incident correlation graph $G_I$ by their reporting services to generate the service correlation graph $G_S$. The edges in $G_I$ are mapped to $G_S$ accordingly. Algorithm~\ref{alg:build_service_correlation_graph} illustrates this process. We first gather all the owning services of the incidents in $G_I$ to $N_{G_S}$. These services are nodes in $G_S$. And then we map the incident links in $G_I$ to the corresponding service links in $G_S$. These links are the edges in $G_S$. This graph helps us  refer to historical anomaly spreading patterns at the service level and helps us predict the root-cause service using the machine learning algorithm.

\begin{algorithm}
\caption{The procedure of building $G_S$}
\begin{flushleft}
    \hspace*{\algorithmicindent} \textbf{Input:} $G_I$ \\
    \hspace*{\algorithmicindent} \textbf{Output:} $G_S$
\end{flushleft}
\begin{algorithmic}[1]
    \Procedure{BuildServiceCorrelationGraph}{}
        \State $N_{G_I}$ $\gets$ nodes in $G_I$
        \State $E_{G_I}$ $\gets$ edges in $G_I$
        \State $N_{G_S} \gets \{\}$
        \State $E_{G_S} \gets \{\}$
        \For {each incident $I_i$ in $N_{G_I}$}
            \State $S$ $\gets$ owning service of $I_i$
            \State add service $S$ to $N_{G_S}$
        \EndFor
        \For {incident link $<I_a, I_b>$ in $E_{G_I}$}
            \State $S_a, S_b$ $\gets$ owning services of $I_a, I_b$
            \State add edge $<S_a, S_b>$ to $E_{G_S}$
        \EndFor
        \State $G_S \gets \{N_{G_S}, E_{G_S}\}$
        \State \Return $G_S$
    \EndProcedure
\end{algorithmic}
\label{alg:build_service_correlation_graph}
\end{algorithm}

\subsection{Training and Predicting}\label{approach_machine_learning}

After the previous steps, although we have reduced the search space of the outage-related incidents and services, the incident correlation graph $G_I$ and the service correlation graph $G_S$ can still be large since the impact of an outage is usually wide-range. Besides, these services may have interdependencies among each other. So engineers still need to dive into the specific services and spend much time for analysis. However, as mentioned in Section~\ref{approach_service_correlation_graph}, the spreading paths of the anomaly among dependent services caused by the same root-cause service have the same pattern, so we can utilize machine learning algorithms to learn the patterns from historical service correlation graphs, and 
to predict the root-cause service automatically. We tried two machine learning algorithms: the SVM (Support Vector Machine) algorithm and the Decision Tree algorithm, which are widely used in a variety of classification tasks.

\textbf{Model Training:} For each outage in the past, we use the methods mentioned above to build its incident correlation graph $G_I$ and service correlation graph $G_S$. The label of each outage is its root-cause service, which is labeled manually after outage diagnosis. And we take the structural information of $G_S$ as a feature vector, which contains two parts: 1) the number of incidents in each service, and 2) the links between these services. 

For the first part, each element represents a service, and its value is the number of incidents included by this service in $G_S$.
For the second part, each element represents a link between two services, and its value is either 0 or 1, indicating the existence of the link in $G_S$. Note that, we have a large number of services, and the links among them form a sparse matrix. This part only includes the service links which occur at least once in the past.

\textbf{Root-cause Service Prediction: } To predict the root-cause service of a new outage, we use the same methods to build its incident correlation graph $G_I$ and service correlation graph $G_S$. We then extract the feature vector from $G_S$ using the same method as in the model training part. Finally, the machine learning model takes the feature vector as the input and predicts the root-cause service as the output.

\section{Experiment}\label{experiment}





In this section, we describe the details of our experiments. We will first present the details of the data we collect for experiments. Then we introduce the techniques we use to implement our approach. Finally, we introduce DeepCT, the state-of-the-art triage approach we use for comparison.

\subsection{Data Collection}

All the data we use in this paper are collected from the production environment of Microsoft Azure. In the IcM system of Microsoft Azure, the data are stored in a distributed NoSQL database. 
We collect 9 months of incident and outage data from the IcM system for training. 
Specifically, we first use the 9 months of data to build the ``incident template -  meta-incident ID'' mapping and the meta-incident ID correlation graph $G_M$. We build the incident correlation graph $G_I$ and the service correlation graph $G_S$ for each outage in the 9 months. We then use the method in Section~\ref{approach_machine_learning} to train the machine learning model.

We collect another 3 months of incident and outage data from the IcM system for testing. The outages in the test dataset happen after the ones in the training dataset. 
We build the incident correlation graph $G_I$ and the service correlation graph $G_S$ for each outage in the 3 months, and use the method in Section~\ref{approach_machine_learning} to predict the root-cause service for each outage.

All these outages involve 225 underlying services in Microsoft Azure and all the data occupies about 133GB of disk space.

\subsection{Implementation}

Our programs are written in \textit{Python3}, a scripting language that is widely used in data mining and machine learning tasks. We also use some third-party \textit{Python3} packages to facilitate our development process.
We use the \textit{Python3} library \textit{Scikit-learn} to build the machine learning models. \textit{Scikit-learn} is an open-source machine learning library that supports supervised and unsupervised learning. 

\subsection{Compared Method}

We compare COT, which is a correlation-based triage approach, with DeepCT~\cite{chen2019continuous}, which is the state-of-the-art triage approach based on text-similarity of incident reports. DeepCT incorporates a novel GRU (Gated Recurrent Unit) model with an attention-based mask strategy and a revised loss function. It can incrementally learn knowledge from discussions and update incident triage results~\cite{chen2019continuous}. DeepCT has demonstrated 
its 
effectiveness on 14 large-scale online service systems.

The data required by DeepCT 
contains three parts: 1) the title and summary of an incident report, 2) the incremental discussions about an incident, and 3) the occurring environment information of an incident. DeepCT uses a CNN-based text encoder to produce feature vectors from the first two parts of data, which are textual. The third part of the data is a ﬁnite set of discrete values, so it uses representation learning~\cite{bengio2013representation} to embed each input-datum value into a ﬁxed-dimension vector~\cite{chen2019continuous}. DeepCT uses these data to train the designed GRU-based model to help predict the root cause of an incident. 

We follow the original paper of DeepCT to implement the CNN-based text encoder, the representation learning model, and the GRU-based model. Since outages are declared from incidents, the three parts of data DeepCT requires are available in outages. So we can easily migrate DeepCT to fit the outage triage problem.
We use the same training and test dataset as COT uses to train and evaluate DeepCT. 

\section{Experimental Results}
\label{evaluation}

In this section, we present the experimental results of our approach and try to answer the following three research questions.

\begin{figure}
    \centering
    \includegraphics[width=0.43\textwidth]{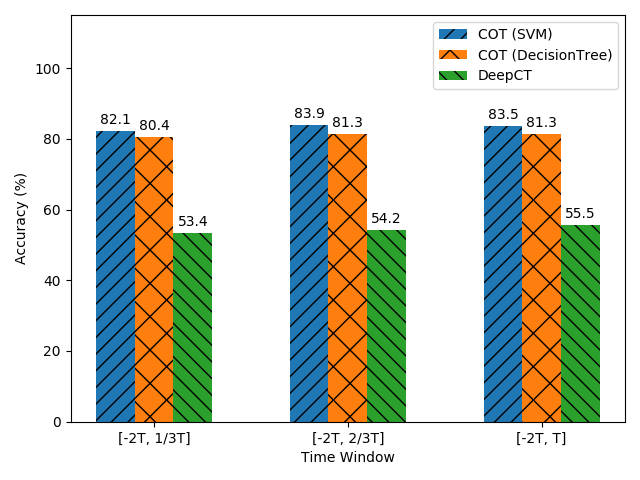}
    \caption{The accuracy of COT and DeepCT in predicting the root-cause service for outages.}
    \label{fig:accuracy-on-service}
\end{figure}

\begin{itemize}
    \item \textbf{RQ1:} How does our approach work in outage triage, comparing to the state-of-the-art triage approach?
    \item \textbf{RQ2:} How is the performance of our approach on different kinds of outages?
    \item \textbf{RQ3:} How is the time efficiency of our approach?
\end{itemize}{}

\subsection{Evaluation Metrics}\label{evaluation_metrics}

The goal of COT is to facilitate the outage triage process, \textit{i.e.}, to help engineers find the correct root-cause service at the early stage of the outage. For each outage, we use COT to predict its root-cause service, compare it to the ground truth, and calculate the accuracy. Besides, to better understand the result, we further inspect its performance on different kinds of outages, \textit{i.e.}, the outages whose root-cause services belong to different service categories.

\subsection{Performance}\label{evaluation_performance}

Figure~\ref{fig:accuracy-on-service} shows the accuracy of COT and DeepCT in predicting the root-cause service 
with different time windows. As it shows, $COT_{SVM}$ outperforms $COT_{DecisionTree}$ slightly at any time window. And these two models outperform DeepCT a lot. The highest accuracy of DeepCT is just 55.5\% and the accuracy of $COT_{SVM}$ is higher than DeepCT by 28.0$\sim$29.7\%. 

As the result shows, the accuracy of $COT_{SVM}$ and $COT_{DecisionTree}$ changes slightly as the time window grows. This tells that the early signals of the outage supply enough information for us to find the root-cause service, and as the time window grows, the accuracy 
does not increase much.

\textbf{Answer to RQ1:} COT can help predict the root-cause service of an outage at its early stage with high accuracy. It outperforms DeepCT by 28.0$\sim$29.7\% in predicting the root-cause service.

\begin{figure}
    \centering
    \includegraphics[width=0.4\textwidth]{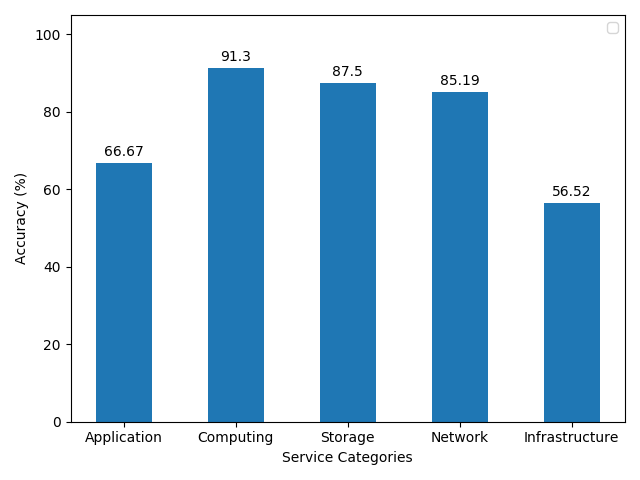}
    \caption{The accuracy of $COT_{DecisionTree}$ for different kinds of outages when setting time window to $[-2T, \frac{1}{3}T]$}
    \label{fig:accuracy-for-each-service-category}
\end{figure}

We use the result of $COT_{DecisionTree}$ with time window $[-2T, \frac{1}{3}T]$ as an example to show the  performance  of  COT on  different kinds of outages
(Figure~\ref{fig:accuracy-for-each-service-category}). Other results have the similar pattern with only slight difference. As Figure~\ref{fig:accuracy-for-each-service-category} shows, for outages whose root-cause services belong to the \textit{Computing}, \textit{Storage}, or \textit{Network} category, the accuracy is relatively high. For outages whose root-cause services belong to the \textit{Application} or \textit{Infrastructure} category, the accuracy is relatively low.

Different from the underlying supporting services, the \textit{Application} services change more frequently, which causes a lot of new signals to emerge. Therefore,  the service correlation graphs for these outages change a lot over time. This phenomenon is natural since the development of user-oriented services should 
satisfy the rapidly changing business requirements. This causes the low accuracy of the $Application$ category.

The low accuracy of the \textit{Infrastructure} category is because the number of outages of this kind is small in our dataset, as the \textit{Infrastructure} services is more robust and fails less frequently. The machine learning algorithm only gains limited knowledge from small samples. Its result should be better as we collect more training samples in the future.

\textbf{Answer to RQ2:}
The  performance  of  COT is different on  different kinds of outages. For the cloud system we studied, COT performs better for outages whose root-cause service is in the \textit{Computing}, \textit{Storage}, or \textit{Network} category. Further research should be taken to tackle the \textit{Application} and \textit{Infrastructure} categories.

\subsection{Time Efficiency}\label{evaluation_time_efficiency}

\begin{figure}
    \centering
    \includegraphics[width=0.45\textwidth]{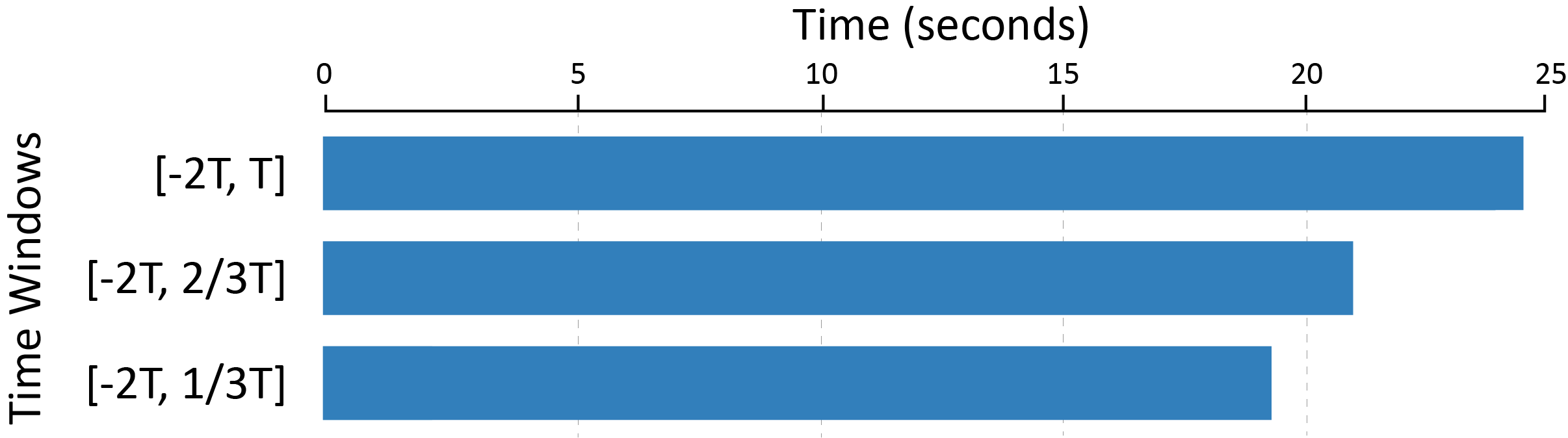}
    \caption{The average time COT spends on predicting the root-cause service for an outage in different time windows.}
    \label{fig:time_efficiency}
\end{figure}

To evaluate the time efficiency of COT, we record the time it spends to predict the root-cause services of outages in different time windows. For the convenience of deployment, we package the COT project as a Docker image, which is a widely used container technique. We deploy it on a machine equipped with a quad-core Intel® Xeon® E5-2673 v4 CPU, and with 16 GB memory. The operating system is Ubuntu 16.04.6 LTS, the Docker version is 18.09.7, and the Python version used in the Docker container is 3.7.

Figure~\ref{fig:time_efficiency} shows the results. For each outage, as the range of the time window grows, the average time of prediction is 19.43$\sim$24.62s. This indicates that COT is efficient for the outage triage task. Our experiment shows that data fetching takes a large proportion of the time. This is because the amount of incidents within the time window is usually large, and the incidents are stored in a large-scale distributed NoSQL database. 
In the future, we can improve the time 
efficiency of COT by caching the incident data and reducing the data fetching time.

\textbf{Answer to RQ3:} The average running time of COT is within one minute, which is efficient for the outage triage task.
\section{Case Study}\label{case_study}

COT is shown to be very helpful for production outage management. In particular, the service correlation graphs built by COT can substantially facilitate the understanding of the complex multi-hop dependencies among services, 
which consequently helps with quick mitigation of the outages. 
In this section, we conduct a case study of two real-world outages ($Outage_A$ and $Outage_B$) in Microsoft Azure to understand how COT works and how COT helps save time in the real-world outage triage scenarios.

\subsection{Two Real-world Cases}

The root cause of $Outage_A$ is some chained reactions of the system followed by the activation of the DDoS defense policy in Microsoft Azure. The DDoS attack is quite frequent in the cloud computing scenario. Thus, cloud computing platforms usually build DDoS defense policy at the network service level to defense such attack. In $Outage_A$, the DDoS defense policy is triggered by some vicious traffic but its reaction is too aggressive, so it also treats other normal network traffic as vicious too. This causes a network configuration server to suffer from network package loss and fail to synchronize the configurations it stores. One important configuration in this server is the mapping from service/component ID to VIPs (Virtual IP Address). Due to the failure of configuration synchronization, this mapping is missing for a DNS service. When other services read this configuration, they cannot gain the exact VIP which matches the DNS service. And following the longest match of the service ID's prefix, what they get is a scope of VIPs belonging to a subnet. This causes abnormal network traffics to the subnet and triggered the DDoS defense policy to protect the subnet, which chooses to drop some network packages targeting at the subnet for some time.

$Outage_B$ is caused by the misconfiguration of some routers. Two engineers intend to modify the TCP MSS (Maximum Segment Size) configuration of two routers. They should have modified this in the network control plane protocol, but they did this in the data plane. The consequence is that every TCP packet received by the router uses intensive CPU time for checking. This causes the CPU usage level to be very high, and the router starts to drop some packages due to the protection policy.

\begin{figure}
\centering
    \begin{subfigure}[b]{0.22\textwidth}
        \includegraphics[width=\textwidth]{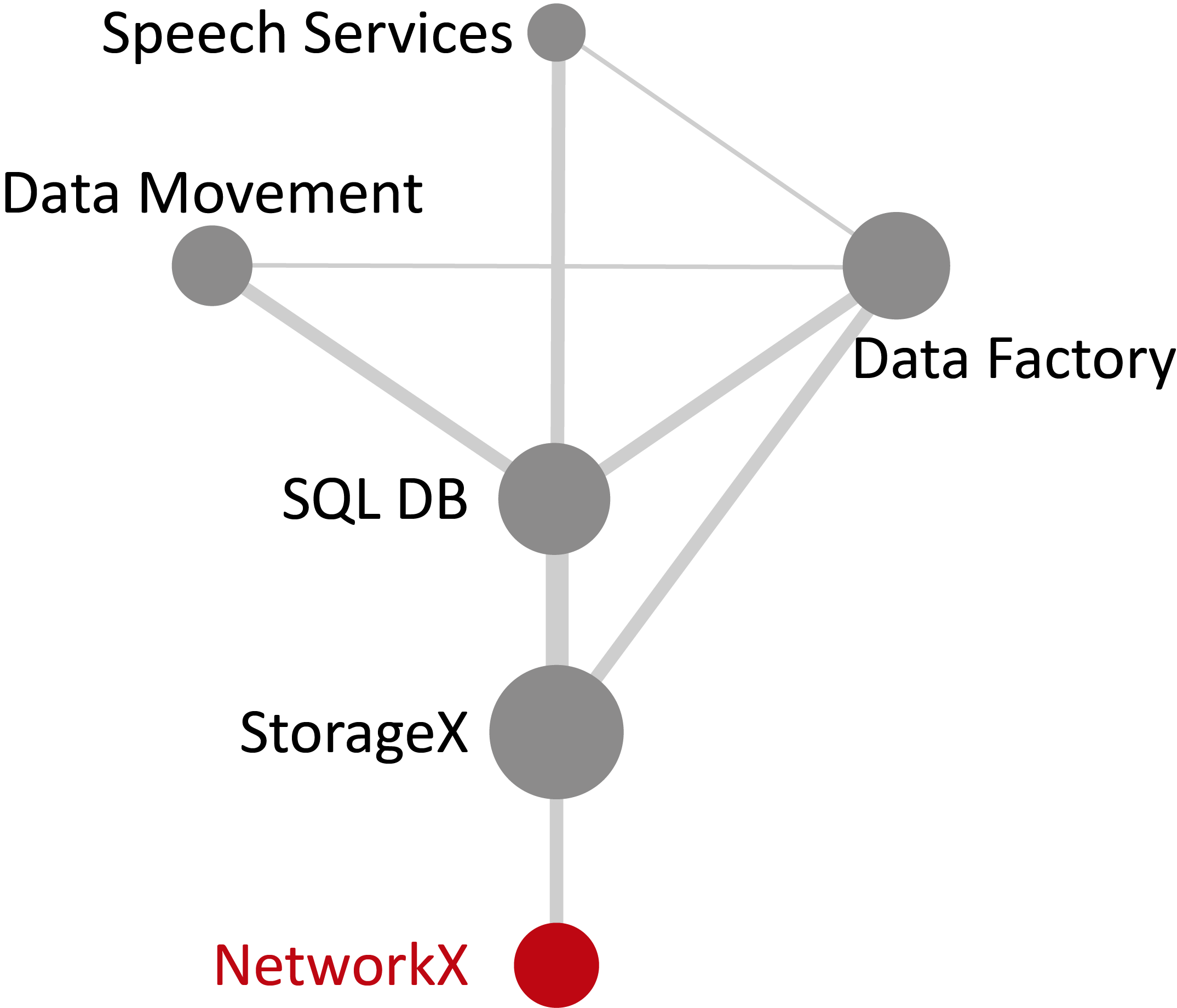}
        \caption{The anomaly spreading pattern among services in $Outage_A$}
        \label{fig:youtube_bad}
    \end{subfigure}\hfill
    \begin{subfigure}[b]{0.22\textwidth}
        \centering
        \includegraphics[width=\textwidth]{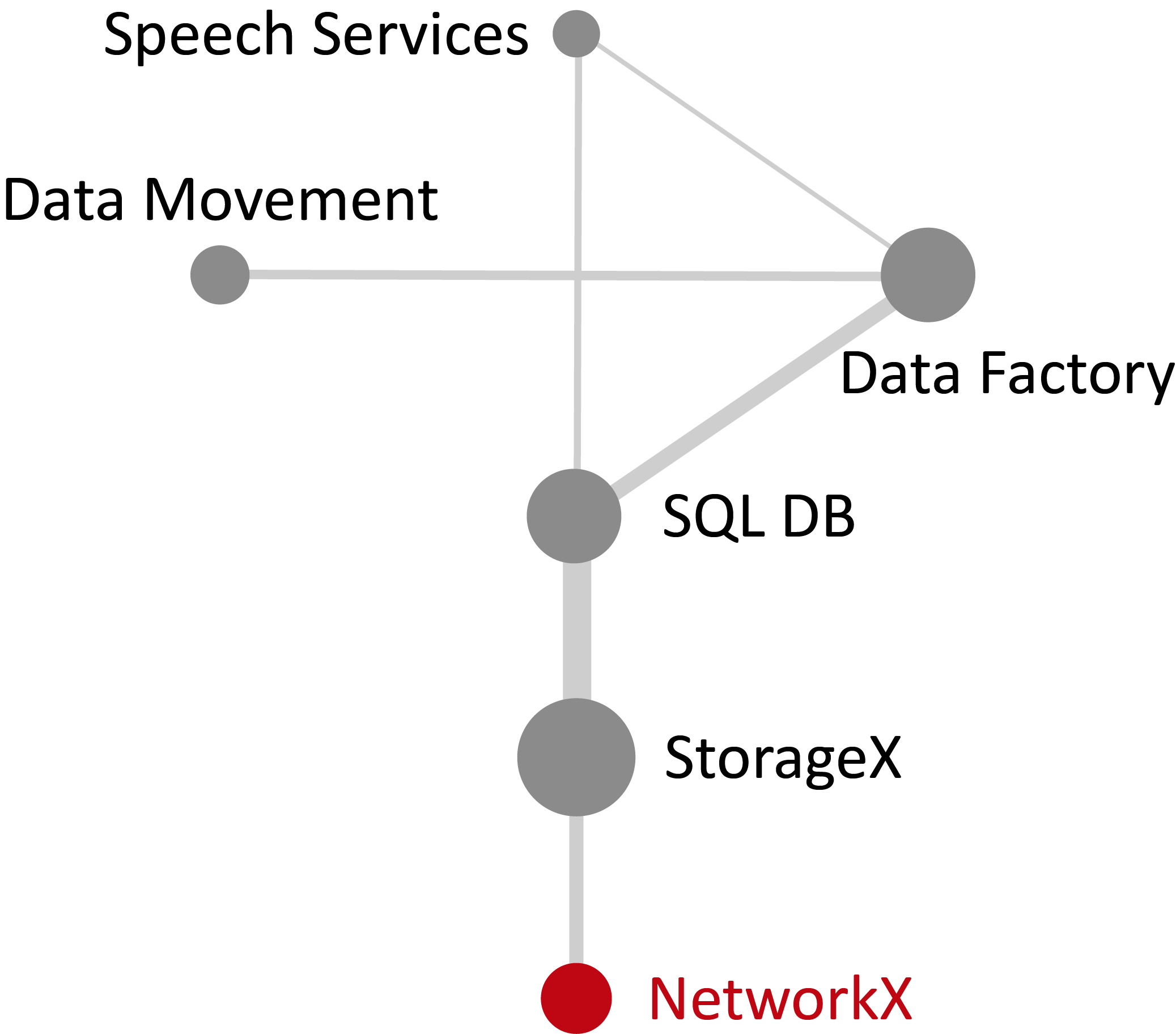}
        \caption{The anomaly spreading pattern among services in $Outage_B$}
        \label{fig:youtube_good}
    \end{subfigure}
\caption{The anomaly spreading pattern among services in $Outage_A$ and $Outage_B$. Each circle in the figure indicates a service in Microsoft Azure. The size of the circle indicates the relative number of incidents reported in that service. Two circles are linked if they are related. Some unimportant services are ignored in this graph for the convenience of explanation.}
\label{fig:case_study_service_correlation_graph}
\end{figure}

The root-cause service is called \texttt{NetworkX} for both outages. But their root-cause bugs are different and are both quite complicated. This causes the monitored metrics in the system to be different and further causes the reporting incidents to be different. For example, the incident reported by \texttt{NetworkX} in $Outage_A$ is a node connectivity failure, and in $Outage_B$ it is a configuration synchronization failure. However, the spreading patterns of the anomaly among services are similar. Figure~\ref{fig:case_study_service_correlation_graph} shows the service correlation graphs of these two outages. 
They all cause the availability of the network services to be low in a subnet. This affects many supporting services which are widely deployed in different regions and sensitive to packet loss and network latency, like the data management services (\texttt{Data Movement} and \texttt{Data Factory}) and the storage services (\texttt{StorageX} and \texttt{SQL DB}). These supporting services further affect the user-oriented services. COT is able to catch the spreading pattern of anomaly starting from the \texttt{NetworkX} service in $Outage_A$, and use this knowledge to predict the root-cause service of $Outage_B$ with high accuracy.

The triage practice for both outages suffers from flooding alarm problems~\cite{chen2020towards}. For example, during the impact of $Outage_B$, there are $K$ incidents reported by 225 services from the affected region of $Outage_B$ in total. Among them, only around 3\% incidents reported from 12 services (5.33\% of 225) are related to $Outage_B$. So it is hard for engineers to find related incidents/services directly in the IcM system. The outage is first declared from a user-oriented service (\texttt{Speech Services}, which is used to convert spoken audio to text), and the triage practice of these two outages is like the case we show earlier in Section~\ref{preliminary_case}, \textit{i.e.,} from the user-oriented service to the real root-cause service, with many reassignments in this triage process. This causes the triage time of these two outages to be very high. COT can save engineers from such a tedious triage process by predicting the root-cause service at the early stage of the outage, quickly and accurately. For example, when setting the time window to $[-2T, \frac{1}{3}T]$, COT can correctly predict the root cause of $Outage_B$ and can help save nearly 80\% of the triage time in theory.

\subsection{Lessons Learned}

We summary the lessons learned from the above two cases  as follows:

\begin{enumerate}
    \item The root-cause bug of an outage is usually very complicated. As these bugs are well repaired after the outage, the same bug may not happen again in the future. This causes the indicators of the impacted service metrics to be different even if the root-cause services of the two outages are the same. So, using only incident content data is not sufficient in the outage triage task.
    \item The traditional outage triage practice is heavily affected by the flooding alarm problem~\cite{chen2020towards} in the IcM. During the impact of an outage, the proportion of related incidents/services is very low. So it is hard for engineers to 
    identify the most important incidents and services. Therefore, in traditional triage practice, manual diagnosis and chained reassignment process are 
    inevitable. 
    \item COT catches the anomaly spreading patterns of different root-cause services from the historical outage diagnosis data. It uses such knowledge to predict the root-cause service for newly occurring outages 
    effectively and efficiently. This helps to save a large amount of outage triage time, and 
    reduce
    the time to mitigate an outage.
\end{enumerate}
\section{Threats to Validity}\label{threats}

COT highly depends on the historical incident correlation data and outage diagnosis data. Currently, we cannot handle zero-day incidents and new anomaly spreading patterns well. But this can be mitigated by continuously updating our model, \textit{i.e.,} regenerate the meta-incident ID correlation graph and retrain the machine learning model regularly. We also 
plan to handle the cases where the services and their dependencies change more frequently (e.g., the $Application$ category in Section~\ref{evaluation_performance}) in our future work.

Another threat comes from our implementation of DeepCT~\cite{chen2019continuous}. We let two members of our team to carefully follow the paper of DeepCT to re-implement it. We test the performance of our implementation 
using the incident triage data of Microsoft Azure, and the results we get are consistent with the 
results described in the original paper, only with small differences. So our implementation should be consistent with the original DeepCT. 

Limited to the data we can obtain, the effectiveness of our approach has only been confirmed in Microsoft Azure. However, as Microsoft Azure is a leading cloud computing platform with a mature IcM system, our experience should be representative and should be applicable to other large-scale cloud computing platforms in the market. Besides, although the data we use cannot be disclosed, the method is reproducible as we have well explained the required details to implement COT in Section~\ref{approach}. In the future, we will try to evaluate COT on other cloud computing platforms to better understand its generality.
\section{Related Work}\label{related}

\subsection{Incident Triage}

Some work focuses on incident triage in large-scale online systems~\cite{chen2019empirical}~\cite{chen2019continuous}~\cite{gu2020efficient}. Chen \textit{et al}.~\cite{chen2019empirical} perform a comprehensive empirical study of incident triage on 20 real-world, large-scale online service systems. Their work studies the status of incident triage from many aspects and explores the practicability of bug triage methods on incident triage. It concludes that traditional bug triage approaches may benefit the incident triage task, but they still need to be further improved to fit the context of incident triage. The work conducted by Chen \textit{et al}.~\cite{chen2019continuous} is the most relevant one to our work. They propose a deep-learning approach named DeepCT to solve the continuous incident triage problem. The authors utilize the historical discussion data of incidents, use the domain-specific text encoding method to extract features from these discussions, and train the deep-learning model. They show the effectiveness of their approach on 14 real-world online service systems.

The above work is different from ours because they mainly focus on incident triage within a single online service, 
while our work focuses on the outage triage problem in the large-scale cloud computing platforms which may involve hundreds of different services of different types.

\subsection{Fault Detection and Localization}


Fault detection and localization in cloud systems have been widely studied in previous work~\cite{Nguyen2013FChain, Chen2014CauseInfer, Sauvanaud2016Anomaly, Thalheim2017Sieve, Marian2018Localizing, Lin2018Microscope}. These methods require constructing the causality graph among components in the system. 
For example, Mariani \textit{et al.}~\cite{Marian2018Localizing} exploit machine learning to detect anomalies in KPIs and exploit the causal relationships among KPIs and centrality indices to identify the causes of the failures. Nguy \textit{ et al.}~\cite{Nguyen2013FChain} propose FChain, which detects anomaly from system metrics (\textit{e.g.}, CPU, memory, network statistics), and locate the cause by using both fault propagation patterns and inter-component dependencies.

However, existing methods are not suitable for outage triage in large-scale production clouds. Firstly, the dependencies among components in cloud computing platforms are incredibly sophisticated. Many services have interdependencies, and many dynamic dependencies are even implicit for engineers. Secondly, when a highly-impactful outage occurs, massive incidents are reported by different services simultaneously. It is also infeasible to apply other methods in our problem settings as they address incidents one by one and greatly suffer from scalability issues. Our method jointly analyzes these incidents. Specifically, COT focuses on the relationship among incidents/services and conducts mitigations altogether to avoid redundant efforts.

\subsection{Bug Triage}

Bug triage has been widely studied in previous work~\cite{lee2017applying,tamrawi2011fuzzy,jeong2009improving,tian2016learning,jonsson2016automated,lin2009empirical,bortis2013porchlight,anvik2011reducing,badashian2016crowdsourced,shokripour2015time,naguib2013bug,shokripour2013so,matter2009assigning,bhattacharya2010fine,wang2014fixercache,linares2012triaging,badashian2015crowdsourced}. Some of the work uses the learning-based approaches. Jonsson \textit{et al}.~\cite{jonsson2016automated} propose an ensemble learning model that combines several classifiers to help assign a bug to the correct development team automatically. Lee \textit{et al}. propose to use a convolutional neural network (CNN) and word embedding techniques to build an automatic bug triager. Some of the work is based on information-retrieval methods. Xia \textit{et al}.~\cite{xia2016improving} propose a method that uses topic modeling to map the bug reports to their corresponding topics. It assigns the bug to the developer by considering the correlation between the developer and the topics of the bug. Hu \textit{et al}.~\cite{hu2014effective} model the relationships among developers, source code components and their associated bugs from historical bug fixing data, and use this knowledge to help assign a bug to the correct developer.

Different from these work that targets at bug triage for traditional software systems, our work focuses on outage triage for large-scale cloud computing platforms, which involve hundreds of distributed services.

\section{Conclusion}\label{conclusion}

In this paper, we conduct the first systematic empirical study on the cross-service outage triage problem in large-scale cloud computing platforms. We also propose COT, a novel data-driven correlation-based outage triage approach. COT mines the anomaly spreading patterns among services caused by different root-cause failures from historical outage diagnosis data. It then uses machine learning algorithms to help predict the root-cause service of a new outage, and to accelerate the outage triage process. We evaluate our approach on the production environment of Microsoft Azure, one of the top cloud providers around the world. 
The data we collected 
occupies about 133GB of disk space and contains records of outages and incidents from a whole year. Experiments show that our model outperforms the state-of-the-art triage approach, which is based on text-similarity, by 28.0$\sim$29.7\% in accuracy, and its overhead is within one minute.

In the future, we intend to 
conduct in-depth research on the cases where services and service dependencies change 
frequently. 
We will also apply our approach to more production cloud systems to have a deeper understanding of the approach's generality. 


\section*{Acknowledgement}
We thank our colleagues in Microsoft Azure 
who developed the incident management system, for their kind help and support in this work. This work was supported by the National Natural Science Foundation of China (Project No. 61672164).
Hongyu Zhang is supported by ARC DP200102940.

\bibliographystyle{ieeetr}
\bibliography{references}
\balance
\end{document}